\documentclass[preprint,aps,showpacs,nofootinbib,preprintnumbers,amsmath,amssymb]{revtex4-1}
\usepackage{}
\usepackage{epsfig}
\usepackage{subfigure}
\usepackage{dcolumn}% Align table columns on decimal point
\usepackage{bm}% bold math
\usepackage[usenames ,dvipsnames]{xcolor}
\usepackage{slashed}
\usepackage{graphicx,color}

\begin{document}
\title{Exploring the simplest purely baryonic decay processes}
\author{C.Q. Geng$^{1}$, Y.K. Hsiao$^{1}$ and Eduardo Rodrigues$^2$}
\affiliation{
$^{1}$Chongqing University of Posts \& Telecommunications, Chongqing,  400065, China\\
%$^{2}$
Physics Division, National Center for Theoretical Sciences, Hsinchu, Taiwan 300\\
%$^{3}$
Department of Physics, National Tsing Hua University, Hsinchu, Taiwan 300\\
$^{2}$Department of Physics, University of Cincinnati, Cincinnati, Ohio 45221, USA
%School of Physics and Astronomy, University of Manchester, Manchester, United Kingdom
}
\date{\today}
\begin{abstract}
Though not considered in general, purely baryonic decays could shed light on
the puzzle of the baryon number asymmetry in the universe by means of a better
understanding of the baryonic nature of our matter world.
As such, they constitute a yet unexplored class of decay processes worth investigating.
We propose to search for purely baryonic decay processes at the LHCb experiment.
No such type of decay has ever been observed.
In particular, we concentrate on the decay
$\Lambda_b^0\to p\bar pn$,  which is the simplest purely
baryonic decay mode, with solely spin-1/2 baryons involved.
We predict its decay branching ratio to be
${\cal B}(\Lambda_b^0\to p\bar pn)=(2.0^{+0.3}_{-0.2})\times 10^{-6}$,
which is sufficiently large to make the decay mode accessible to LHCb.
Our study can be extended to other purely baryonic decays such as
$\Lambda_b^0\to p\bar p \Lambda$ and $\Lambda_b^0\to \Lambda\bar \Lambda\Lambda$,
as well as to similar decays of antitriplet $b$ baryons such as $\Xi_b^{0,-}$.
\end{abstract}
%\pacs{}

\maketitle
\section{introduction}
It is well known that every (anti)baryon except the (anti)proton decays 
to a lighter (anti)baryon, such as the beta decay of the neutron,
$n\to p e^-\bar{\nu}_e$, which is the simplest baryonic decay. 
However, up to now, no {\em purely} baryonic decay process, with only baryons involved,
has yet been observed~\cite{pdg}.
By virtue of the baryon number conservation, in order to have a purely baryonic decay, 
at least three baryons in the final state are needed, e.g.,
${\bf B}_h\to{\bf B}_{l_1} \bar {\bf B}_{l_2} {\bf B}_{l_3}$, 
where $h$ and $l_i$ represent heavy and light spin-1/2 baryons, respectively.
It is easy to show that the simplest, and lightest, possible purely baryonic decay process is 
$\Lambda_b^0\to p\bar p n$ without breaking any known conservation law.
Other examples of such decays are  
$\Lambda_b^0\to p\bar p \Lambda$ and $\Lambda_b^0\to \Lambda\bar \Lambda\Lambda$ as well as the
corresponding $\Xi_b^{0,-}$ decays.
Since baryons are the main constituents of our matter world,
their production and decay mechanisms should all be explored.

Being one of the three conditions for the baryogenesis to explain the puzzle of
the matter and antimatter asymmetry in the universe,
$CP$ violation has been a primary topic of study at the $B$ factories and at the LHCb experiment, among others.
However, the unique physical $CP$ phase in the 
Cabibbo-Kobayashi-Maskawa (CKM) quark mixing matrix~\cite{CKM}
of the Standard Model (SM) is not sufficient to 
solve the mystery, leaving room for new physics in the formation of the matter world.
On the other hand, for a direct connection to the baryonic contents of the universe,
one expects to observe  $CP$ violation in purely baryonic processes.
Theoretically and experimentally, though, the latter has,
to our knowledge, never been studied before.
Clearly, 
it is interesting to discuss $CP$ violating rate asymmetries as well as time-reversal violating
spin involved triple correlations due to the rich spin structures in these purely baryonic decays,
to test the SM and search for new physics manifestations.

In the SM, 
the decay $\Lambda_b^0\to p\bar p n$, with the tree-level dominated contribution 
through the $V\!\!-\!\!A$ quark currents, can be factorized as a color-allowed process, 
which is insensitive to nonfactorizable effects, where
the required matrix elements of the $\Lambda_b^0\to p$ transition and 
the recoiled $\bar p n$ pair have been well studied. 
A reliable prediction of the branching ratio is hence expected, which should be 
as large as those in the tree-level $B$ decays such as $\bar B^0\to \pi^+\pi^-$.
In addition, the threshold effect 
around the $\bar p n$ invariant mass spectrum, measured as a salient feature
in three-body baryonic $B$ decays, could also enhance the contribution.

It is interesting to note that in this simplest purely baryonic decay $\Lambda_b^0\to p\bar p n$,
a possible intermediate resonant state such as $D_s^- (\to \bar p n) p$ is rather suppressed,
which is estimated to have
${\cal B}(\Lambda_b\to p (D^-_s\to) n\bar p)\simeq 
{\cal B}(\Lambda_b\to p D^-_s){\cal B}(D^-_s\to n\bar p)\simeq 
2\times 10^{-8}$~\cite{Athar:2008ug,Hsiao:2015cda}.
While the threshold effect, receiving the dominant contribution from the threshold of 
$m_{\bf B\bar B'}\simeq m_{\bf B}+m_{\bf \bar B'}$, has
been commonly observed in the baryon pair production~\cite{Chua:2001xn,Geng:2006yk},
the meson resonances or the final state interaction
due to the multiparticle exchange, 
which deviates the baryon pair production from the threshold, should be suppressed.

In this article, we concentrate on this simplest purely baryonic decay 
$\Lambda_b^0\to p\bar p n$. We will give the theoretical estimation of its decay
branching ratio, and stimulate a possible measurement by the LHCb Collaboration.
Possible $CP$ and $T$ violating effects in purely baryonic decays are also discussed.

\section{Theoretical predictions}
%Formalism}
%=======================
\begin{figure}[t!]
\centering
\includegraphics[width=2.65in]{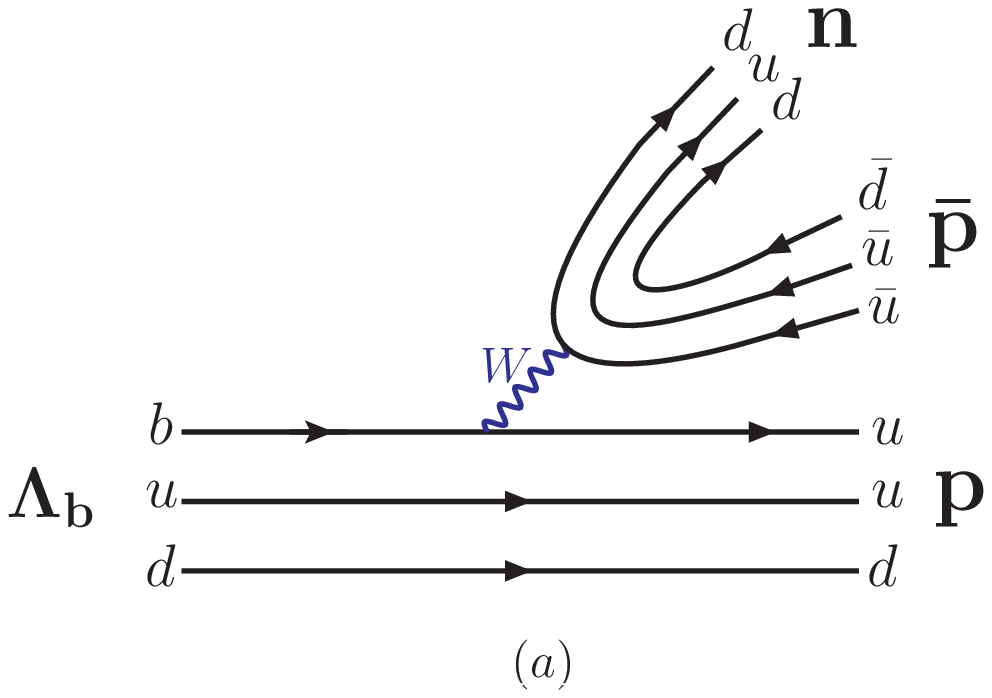}
\includegraphics[width=2.5in]{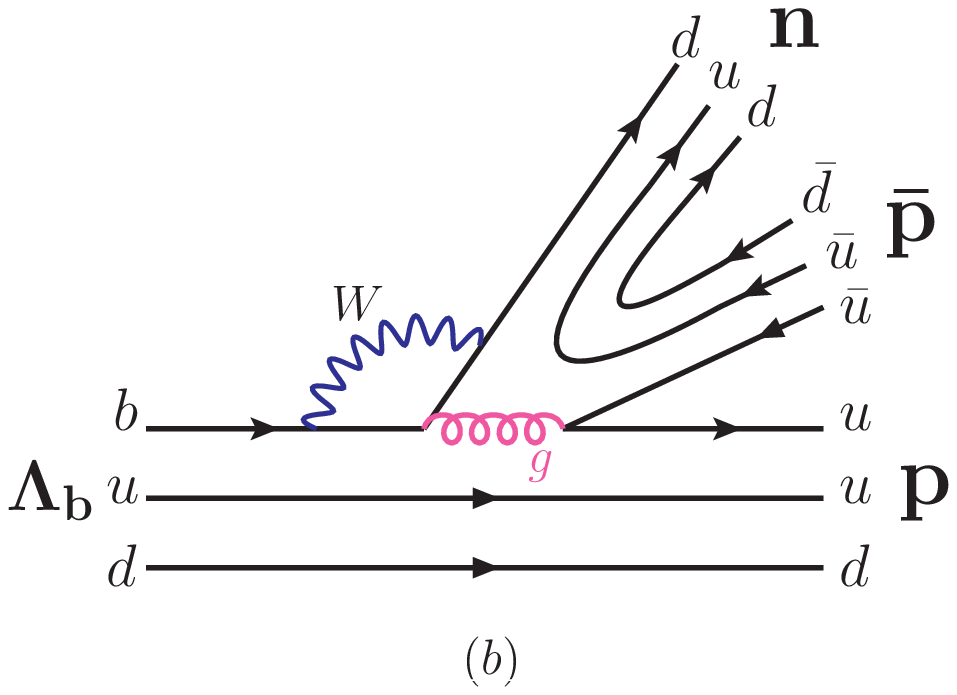}
\includegraphics[width=2.5in]{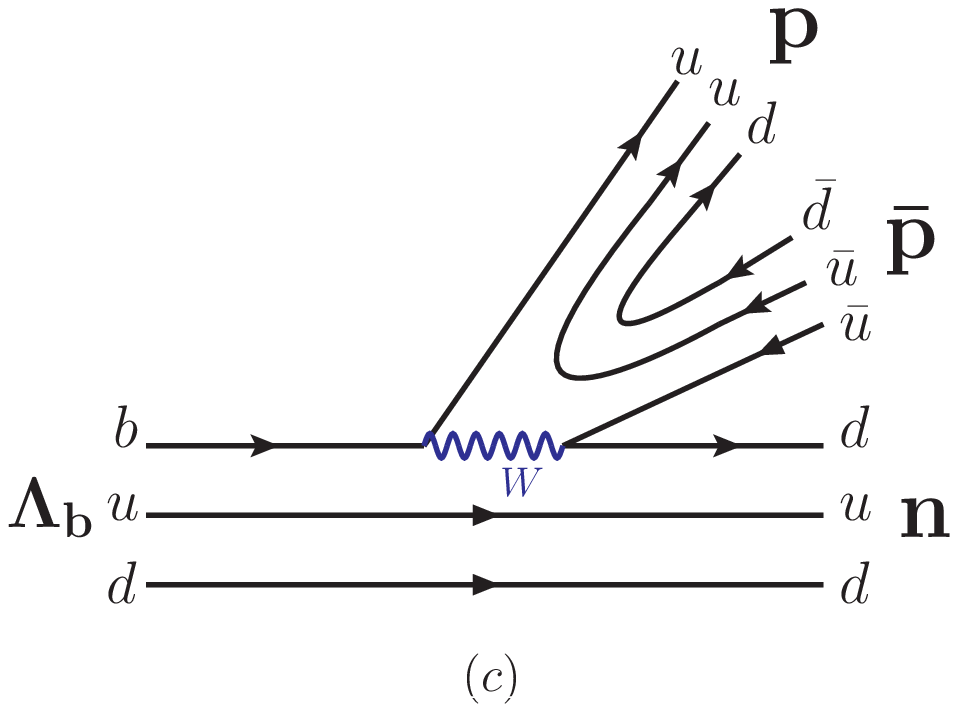}
\includegraphics[width=2.65in]{Lbtoppbarnt2.eps}
\caption{(a), (c) Tree-level and (b), (d) penguin-level Feynman diagrams
         contributing to the $\Lambda_b^0\to p\bar p n$ decay.}\label{dia}
\end{figure}
%======================
In terms of the effective Hamiltonian at the quark level for
the charmless $b\to u\bar ud$ transition in the SM, 
the amplitude for $\Lambda_b^0\to p\bar p n$ in the factorization approach 
can be written as
\begin{eqnarray}\label{amp1}
{\cal A}(\Lambda_b^0\to p\bar p n)\simeq \frac{G_F}{\sqrt 2}V_{ub}V_{ud}^* a_1
\langle  n\bar p|(\bar d u)_{V-A}|0\rangle
\langle p|(\bar u b)_{V-A}|\Lambda_b^0\rangle\,,
\end{eqnarray}  
where
$G_F$ is the Fermi constant; $V_{ij}$ are the CKM matrix elements;
$(\bar q_1 q_2)_{V(A)}$ stands for $\bar q_1 \gamma_\mu(\gamma_5) q_2$; and
$a_i=c^{\rm eff}_i+c^{\rm eff}_{i\pm1}/N_c$ for $i=$odd (even) 
with  the effective Wilson coefficients $c^{\rm eff}_i$
defined in Ref.~\cite{ali} and the color number $N_c$, which 
floats from 2 to $\infty$ to estimate the nonfactorizable effects
in the generalized version of the factorization approach.
The amplitude in Eq.~(\ref{amp1}) is dominated by 
the tree contribution from Fig.~\ref{dia}(a),
while that from Fig.~\ref{dia}(c) is primarily the nonfactorizable effect
with its size proportional to $a_2$.  
Since the global fittings for $a_2$ indicate
a universal value of ${\cal O}(0.2-0.3)$~\cite{Neubert:2001sj,Hsiao:2015cda,Hsiao:2016amt}, 
$a_2$ is within the uncertainty of $a_1$.
On the other hand,
the penguin contributions in Figs.~\ref{dia}(b) and \ref{dia}(d)  have been neglected 
due to the suppressed values of
$|(V_{tb}V_{td}^*)/(V_{ub}V_{ud}^*)(a_i/a_1)|^2 \le 0.02$ ($i>2$).

In Eq.~(\ref{amp1}), the matrix elements for the baryon pair production 
are well defined, given by
\begin{eqnarray}\label{timelikeF}
\langle n\bar p|\bar d\gamma_\mu u|0\rangle
%&=&
%\bar u\bigg\{F_1\gamma_\mu+\frac{F_2}{m_{\bf B}+m_{\bf \bar B'}}i\sigma_{\mu\nu}q_\mu\bigg\}v\;,\nonumber\\
&=& \bar u_n\bigg\{[F_1+F_2]\gamma_\mu
+\frac{F_2}{m_n+m_{\bar p}}(p_{\bar p}-p_n)_\mu\bigg\}v_{\bar p}\;,\nonumber\\
\langle  n\bar p|\bar d\gamma_\mu \gamma_5 u|0\rangle&=&
\bar u_n\bigg\{g_A\gamma_\mu+\frac{h_A}{m_n+m_{\bar p}}q_\mu\bigg\}\gamma_5 v_{\bar p}\,,
\end{eqnarray}
where $q=p_n+p_{\bar p}$ %=p_{{\cal B}_b}-p_{\bf B_n}$ 
is the momentum transfer,
$F_{1,2}$, $g_A$ and $h_A$ are the timelike baryonic form factors, and
$u_n$($v_{\bar p}$) is the neutron (antiproton) spinor. 
On the other hand, the matrix elements of the $\Lambda_b\to p$ baryon
transition in Eq. (\ref{amp1}) have the general forms:
\begin{eqnarray}
&&\langle p|\bar u \gamma_\mu b|\Lambda_b\rangle=
\bar u_p\left[f_1\gamma_\mu+\frac{f_2}{m_{\Lambda_b}}i\sigma_{\mu\nu}q^\nu+
\frac{f_3}{m_{\Lambda_b}}q_\mu\right] u_{\Lambda_b}\,,\nonumber\\
&&\langle p|\bar u \gamma_\mu\gamma_5 b|\Lambda_b\rangle=
\bar u_p\left[g_1\gamma_\mu+\frac{g_2}{m_{\Lambda_b}}i\sigma_{\mu\nu}q^\nu+
\frac{g_3}{m_{\Lambda_b}}q_\mu\right]\gamma_5 u_{\Lambda_b}\,,
\end{eqnarray}
where $f_j$ ($g_j$) ($j=1,2,3,$)  are the form factors, with
$f_1=g_1$ and $f_{2,3}=g_{2,3}=0$ resulting from
the SU(3) flavor and SU(2) spin symmetries~\cite{Hsiao:2015cda}, which 
agree with the results based on the heavy-quark and large-energy symmetries in Ref.~\cite{CF}.

%\section{Numerical Results and Discussions}
For the numerical analysis, the theoretical inputs of the CKM matrix elements 
in the Wolfenstein parametrization are given by~\cite{pdg}
\begin{eqnarray}\label{para}
V_{ub}=A\lambda^3 (\rho-i\eta)\,,\;\; V_{ud}=1-\lambda^2/2\,,
\end{eqnarray}
with $(\lambda,\,A,\,\rho,\,\eta)=(0.225,\,0.814,\,0.120\pm 0.022,\,0.362\pm 0.013)$.
We adopt 
$(c^{\rm eff}_1,\,c^{\rm eff}_2)=(1.168,\,-0.365)$ in Ref.~\cite{ali}, and
obtain $a_1=1.05^{+0.12}_{-0.06}$.
%
%For $F_1$ ($g_A$) and $f_1$ ($g_1$),
% are associated with
%the studies of baryonic $B$ decays and $b$-baryon decays, respectively, 
%In terms of the momentum dependences of
For the timelike baryonic form factors, it is adopted that
$F_1(g_A)={C_{F_1(g_A)}}/{t^2}[\text{ln}({t}/{\Lambda_0^2})]^{-\gamma}$,
$h_A=C_{h_A}/t^2$, and $F_2=F_1/(t\text{ln}[t/\Lambda_0^2])$
with $\gamma=2.148$, $\Lambda_0=0.3$ GeV and $t\equiv q^2$
in pQCD counting rules~\cite{Brodsky1,Hsiao:2014zza,F2}.
The form factors at the timelike region may possess the strong phase 
with the analytical continuation to the spacelike region~\cite{Brodsky:2003gs}, 
since it can be derived as an overall factor to all subprocesses in Fig.~\ref{dia}.
This strong phase, in fact has no effect and is neglected in our paper.
The form factors with the asymptotic behaviors in pQCD counting rules
have been justified to agree with the $e^+ e^-\to p\bar p(n\bar n)$ data
at $t=(4-10)$ GeV$^2$~\cite{Chua:2001xn,Geng:2006yk}.
Besides, they have been used to explain the branching ratios, and the so-called threshold effect 
in the baryon pair invariant mass spectra, which
presents the peak around the threshold of $t\simeq 4$ GeV$^2$ and gradually 
turns out to be flat around $t=10-16$ GeV$^2$~\cite{ppD,Hou:2000bz,Hsiao:2016amt}, 
being observed as the common feature in the three-body baryonic $B$ decays.
For the $\Lambda_b\to p$ transition,
the $f_1(g_1)$ is presented as the double-pole momentum dependences:
$f_1(g_1)=C_{f_1(g_1)}/(1-t/m_{\Lambda_b}^2)^2$~\cite{Hsiao:2014mua}.
Since the form factors are associated with
the studies of baryonic $B$ decays and $b$-baryon decays,
we use the numerical results 
in Refs.~\cite{Hsiao:2014zza,Hsiao:2014mua,Wei:2009np,ppD,Hsiao:2015cda,CF}
to give 
$(C_{F_1},C_{g_A}, C_{h_A})=(196.1\pm 37.6,101.0\pm 37.6,-4.5\pm 2.2)$~GeV$^4$, and $C_{f_1}=C_{g_1}=0.136\pm 0.009$.
%where $C_T$ is consistent with the value of  $0.14\pm 0.03$ 
%obtained in the  lght-cone sum rules approach~\cite{CF}, as well as 
%in other theoretical calculations~\cite{Wei:2009np}.
%

With all theoretical inputs, we find
\begin{eqnarray}
{\cal B}(\Lambda_b^0\to p\bar p n)=(2.0^{+0.3}_{-0.2}\pm0.1\pm 0.1)\times 10^{-6}\,,
%{\cal B}(\Lambda_b^0\to p\bar p n)=(2.00^{+0.36}_{-0.20})\times 10^{-6}\,,
\end{eqnarray}
where the errors come from the form factors, 
the nonfactorizable effects, and 
the CKM matrix elements, respectively.
By combining the uncertainties, we obtain the first prediction,
${\cal B}(\Lambda_b^0\to p\bar p n)=(2.0^{+0.3}_{-0.2})\times 10^{-6}$,
on the branching ratio of this purely baryonic decay,
which is sizable and comparable to the branching ratios of
other baryonic $B$ decays observed at the $B$ factories and LHCb.

\begin{figure}[t!]
\centering
\includegraphics[width=2.65in]{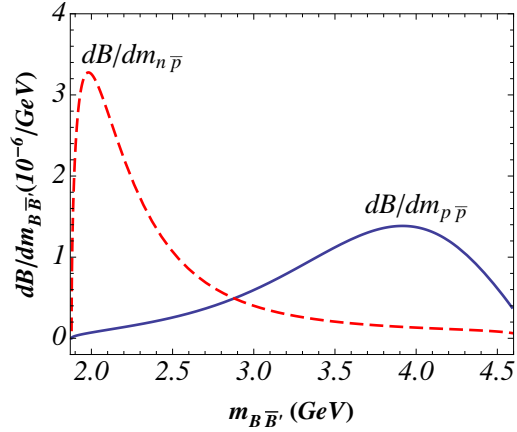}
\caption{The dibaryon invariant mass spectra for $\Lambda_b^0\to p\bar p n$.}\label{mBB}
\end{figure}

Figure~\ref{mBB} displays the invariant mass spectra of $m_{n\bar{p}}$ and $m_{p\bar{p}}$.
The neutron-antiproton invariant mass spectrum 
in $\Lambda_b^0\to p\bar p n$ presents the threshold effect
due to the form of $1/t^n$ %in Eq.~(\ref{timelikeF2})
for the $n\bar p$-pair production, 
which is similar to the peak around the threshold area of 
$m_{\bf B\bar B'}\simeq m_{\bf B}+m_{\bf \bar B'}$ 
that enhances the branching ratios 
in the three-body baryonic $B\to {\bf B\bar B'}M$ decays.
On the other hand, the $m_{p\bar p}$ distribution is in accordance with
the fact that the proton and antiproton are not pair produced.
Note that the spectra in Fig.~\ref{mBB} are partly a consequence of ignoring
the contributions from the diagrams in Figs.~\ref{dia}(c) and~\ref{dia}(d).
Future measurements of dibaryon spectra should be able to test 
our assumptions, in particular
 the factorization approach. 
%
%=======================
%======================

Since purely baryonic decays are
directly connected to the baryonic contents of the universe,
it is worthwhile to have systematic investigations
of their decay branching ratios and direct $CP$ violating rate asymmetries
as well as the possible $T$-odd triple correlations.
Besides the simplest mode $\Lambda_b^0\to p \bar p n$,
example decays are 
$\Lambda_b^0\to p\bar p\Lambda$ and $\Lambda_b^0\to \Lambda\bar \Lambda\Lambda$ decays,
and decays of other antitriplet $b$ baryons such as $\Xi_b^{0,-}$.
The direct $CP$ violating rate asymmetry can be defined by
\begin{eqnarray}\label{acp1}
{\cal A}_{CP}=\frac{\Gamma( 
{\bf B}_h\to{\bf B}_{l_1} \bar {\bf B}_{l_2} {\bf B}_{l_3})
-\Gamma( 
{\bf\bar B}_h\to{\bf \bar B}_{l_1}  {\bf  B}_{l_2} {\bf\bar B}_{l_3})}
{\Gamma( 
{\bf B}_h\to{\bf B}_{l_1} \bar {\bf B}_{l_2} {\bf B}_{l_3})
+\Gamma( 
{\bf\bar B}_h\to{\bf \bar B}_{l_1}  {\bf  B}_{l_2} {\bf\bar B}_{l_3})
}\;.
\end{eqnarray}
If both weak ($\gamma$) and strong ($\delta$) phases
are nonvanishing, one has that
\begin{eqnarray}\label{acp2}
{\cal A}_{CP}\propto \sin\gamma\sin\delta\;.
\end{eqnarray}
For $T$ violation in purely baryonic decays, 
we can examine the triple product correlations  of the $T$-odd form
$\vec v_1\cdot(\vec v_2 \times \vec v_3)$ where $\vec v_i$ 
 are spins or momenta of the baryons.
  Explicitly, for this class of decays
 ${\bf B}_h\to{\bf B}_{l_1} \bar {\bf B}_{l_2} {\bf B}_{l_3}$, 
 there are many $T$-odd correlations of 
  $\vec s_{{\bf B}_j} \cdot (\vec p_{{\bf B}_k} \times \vec p_{{\bf B}_l})$
  and $\vec p_{{\bf B}_j} \cdot (\vec s_{{\bf B}_k} \times \vec s_{{\bf B}_l})$,
  corresponding to one and two spins, respectively,
  due to the rich spin structures of the baryons.
     The asymmetry depending on these correlations is defined by
 \begin{eqnarray}\label{at1}
A_T\equiv\frac{\Gamma(\vec v_1\cdot(\vec v_2 \times \vec v_3)>0)
-\Gamma(\vec v_1\cdot(\vec v_2 \times \vec v_3)<0)}
{\Gamma(\vec v_1\cdot(\vec v_2 \times \vec v_3)>0)
+\Gamma(\vec v_1\cdot(\vec v_2 \times \vec v_3)<0)}\,,
\end{eqnarray}
which results in
\begin{eqnarray}\label{at2} A_T\propto
\sin(\gamma+\delta).
\end{eqnarray}
Note that $A_T$ may not indicate the real $CP$ violating effect
 in the gauge theory.
To obtain the true
effect, one can construct the asymmetry  by using
\begin{eqnarray}
{\cal A}_T\equiv\frac{1}{2}(A_T-\bar A_T)\;,
\label{at12}
\end{eqnarray}
where $\bar A_T$ is measured in the $CP$-conjugate decay process,
and  consequently one finds that
\begin{eqnarray}\label{at3}
{\cal A}_T\propto\sin\gamma\cos\delta,
\end{eqnarray}
which is in general nonzero
%where ${\cal A}_T$ disappears 
as long as $\gamma\neq 0$, no matter whether
the strong phase $\delta$ exists.
We remark that, for the baryonic decays with mesons,
such as $\Lambda_b\to p K^{(*)-}$, $\bar B^0\to \Lambda\bar p\pi^+$, and $B^-\to p\bar p K^{(*)-}$,
their branching ratios, and the $CP$ and $T$ violating asymmetries have been well 
studied~\cite{Bensalem,Geng:2005wt,Hsiao:2014mua}.
In contrast, the purely baryonic decays are conceptually new species
and have not been well explored yet.
In the SM, 
the direct $CP$ violating asymmetry in Eq.~(\ref{acp1}) and the $T$ violating asymmetry in Eq.~(\ref{at12})
are estimated to be both around -4\%.
As some new $CP$ violating mechanism is anticipated, these asymmetries could be sensitive to it.
For example, 
the asymmetries can be enhanced in the $CP$ violating models with  charged Higgs and gauge bosons.

%\cite{Bensalem:2002pz}
%\bibitem{Bensalem:2002pz} 
%``T violating triple product correlations in charmless Lambda(b) decays,''
%W.~Bensalem, A.~Datta and D.~London, Phys.\ Lett.\ B {\bf 538}, 309 (2002) [hep-ph/0205009].

%\cite{Bensalem:2002ys}
%\bibitem{Bensalem:2002ys} 
%``New physics effects on triple product correlations in Lambda(b) decays,''
%W.~Bensalem, A.~Datta and D.~London, Phys.\ Rev.\ D {\bf 66}, 094004 (2002) [hep-ph/0208054].

\section{Experimental considerations}
Experimentally speaking, the presence of a neutron in the final state gives rise
to a signature resembling that of a semileptonic decay. The LHCb experiment has
already published studies of such topologies with two charged tracks and a particle
escaping detection, for example $\Lambda_b^0\to p \mu \bar \nu_\mu$~\cite{LHCb_Vub}.
The analysis can be seen as a proof of concept for the study of
$\Lambda_b^0\to p\bar p n$.

The observation of the $\Lambda_b^0\to p\bar p n$ signal is nevertheless rather
challenging given a branching ratio significantly lower than that of a typical
semileptonic decay. Decays with a $p \bar{p}$ pair in the final state and
extra invisible or nonreconstructed particles are potentially dangerous sources
of background, with branching ratios in the same range -- $10^{-6}$ -- as the signal.
%It is promising to observe this simplest purely baryonic decay.

\begin{figure}[htbp!]
\centering
\includegraphics[width=4.0in]{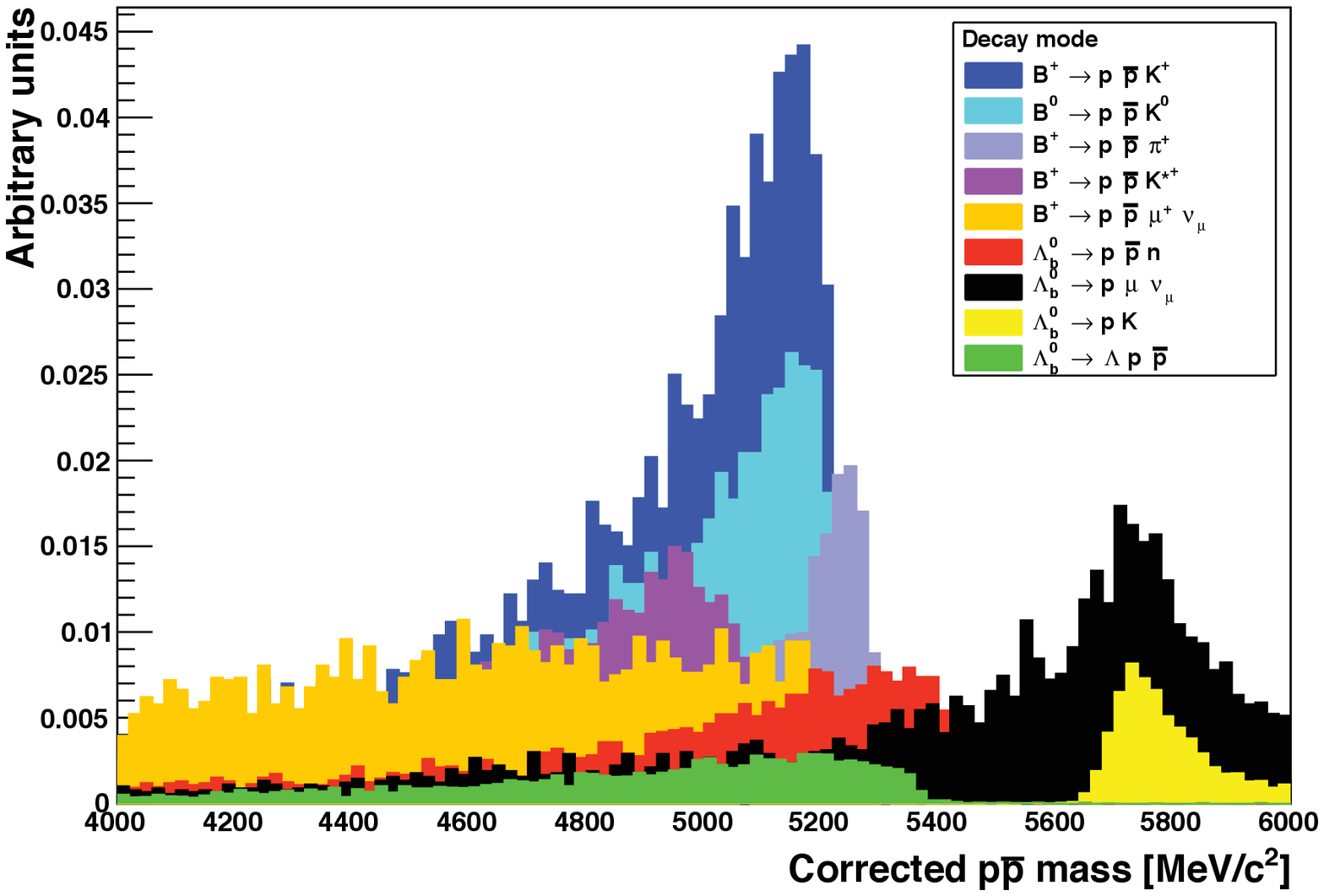}
\includegraphics[width=4.0in]{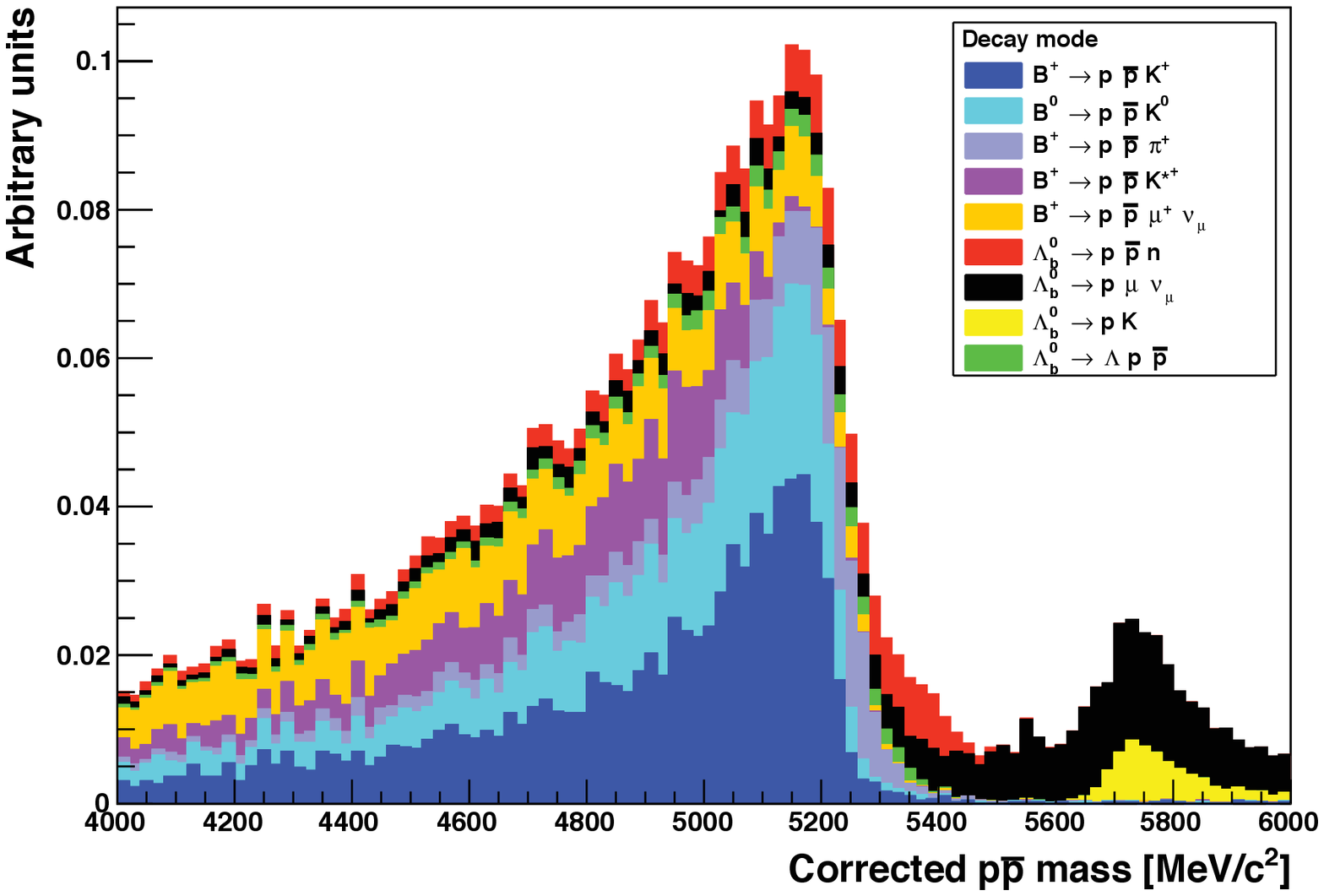}
\caption{Distributions of the $p\bar{p}$ pair corrected mass for the signal
$\Lambda_b^0\to p\bar pn$ and various sources of background
resulting from a toy simulation study.
The top part presents all distributions normalized to the unit area,
whereas the bottom part stacks all contributions so as to give a more realistic
picture of the kind of spectrum at hand.
Refer to the text for further details.}\label{PPbarCorrM}
\end{figure}

The $\Lambda_b^0$ candidate can be reconstructed using the so-called
\textit{corrected mass}~\cite{Abe:1997sb} defined by
$$
m_{\rm corr} = \sqrt{m_{p\bar{p}}^2 + p_\perp^2} + p_\perp\;,
$$
where $m_{p\bar{p}}$ is the invariant mass of the $p\bar{p}$ pair and
$p_\perp$ its momentum transverse to the $\Lambda_b^0$ direction of flight.
Figure~\ref{PPbarCorrM} shows the distributions of the corrected $p\bar{p}$ mass
for the signal and a few typical background decay modes resulting from a toy
simulation study. Care has been taken to smear the momentum resolution of tracks
according to the average resolution published by the LHCb experiment.
Also the relative contributions have been scaled taking into account the
experimental branching ratios (reasonable assumptions on the branching ratio
were made when the decay mode has not yet been observed)
and typical misidentification rates in LHCb~\cite{LHCbDetPerf}.

The bottom figure gives in particular the sum over all contributions.
The signal appears as a shoulder around the region
$5300-5500\mathrm{\,Me\kern -0.1em V\!/}c^2$.
It is evident that isolation requirements similar to those implemented e.g.,
in Ref.~\cite{LHCb_Vub}, need to be exploited in order to control the decays to
$p \bar{p} X$ final states, where $X$ represents one or several charged and/or neutral
particles. An interesting alternative is the identification of the signal neutron
with the calorimeter of the LHCb experiment. 
The authors are aware that no publication from
LHCb has ever studied neutrons, which makes the route challenging. 
These preliminary studies do indicate, though, that the observation of 
$\Lambda_b^0\to p\bar pn$ at LHCb is promising.

For $CP$ and/or $T$ violation, although the current sensitivity of  LHCb is unable to reach 
the level predicted in the SM, it is still worthwhile to explore the $CP$/$T$ violating asymmetries
of fully reconstructed baryonic decays --  $\Lambda_b^0\to p\bar p \Lambda$ is a prominent
example -- as they could be large in the new $CP$ violating models. 

%=======================
%======================

\section{Conclusions}
Since $CP$ violation in fully baryonic decay processes 
is directly related to the matter and antimatter asymmetry of the universe,
we have studied the simplest case of $\Lambda_b^0\to p\bar pn$
to investigate its accessibility at the LHCb experiment.
With the predicted ${\cal B}(\Lambda_b^0\to p\bar pn)=(2.0^{+0.3}_{-0.2})\times 10^{-6}$,
this decay can be the new frontier to test the SM and search for new physics.
One can, and should, study other purely baryonic decay modes:
$\Lambda_b^0\to p\bar p \Lambda$, $\Lambda_b^0\to \Lambda\bar \Lambda\Lambda$ and 
other similar decays of antitriplet $b$ baryon such as $\Xi_b^{0,-}$.

\section*{ACKNOWLEDGMENTS}
E. R. wishes to thank the National Center for Theoretical Sciences at Hsinchu for its warm hospitality.
The work was supported in part by National Center for Theoretical Sciences, National Science
Council (NSC-101-2112-M-007-006-MY3), MoST (MoST-104-2112-M-007-003-MY3) and National Tsing Hua
University (104N2724E1).

\end{document}